\documentclass{PoS}
\newcommand{\beq}{\begin{equation}}
\newcommand{\eeq}{\end{equation}}
\newcommand{\bea}{\begin{eqnarray}}
\newcommand{\eea}{\end{eqnarray}}

\newcommand\trt{\frac14\textrm{Tr}}
\newcommand\avr[1]{\left\langle{#1}\right\rangle}

\title{Freeze-out parameters from continuum extrapolated lattice data}

\ShortTitle{Freeze-out parameters from continuum extrapolation}

\author{\speaker{Sz. Bors\'anyi}$^1$,
Z. Fodor$^{1,2,3}$, S. D. Katz$^{2.4}$, S. Krieg$^{1,3}$, C. Ratti$^{5}$, K. K. Szab\'o$^{1}$
\\
$^1$ \small{\it Department of Physics, Wuppertal University, Gaussstr. 20, D-42119
 Wuppertal, Germany}\\
$^2$ \small{\it Inst. for Theoretical Physics, E\"otv\"os University,}\\
\small{\it P\'azm\'any P. s\'et\'any 1/A, H-1117 Budapest, Hungary}\\
$^3$ \small{\it J\"ulich Supercomputing Centre, Forschungszentrum J\"ulich, D-52425
J\"ulich, Germany}\\
$^4$ \small{\it MTA-ELTE "Lend\"ulet" Lattice Gauge Theory Research Group,}\\
\small{\it P\'azm\'any P. s\'et\'any 1/A, H-1117 Budapest, Hungary}\\
$^5$ \small{\it Dip. di Fisica, Universit\`a di Torino and INFN, Sezione di Torino}\\
\small{\it via Giuria 1, I-10125 Torino, Italy}\\
        E-mail: \email{borsanyi@uni-wuppertal.de}}

\abstract{
We present continuum extrapolated lattice results for the higher order
fluctuations of conserved charges in high temperature Quantum Chromodynamics.
Through the matching of the grand canonical ensemble on the lattice to
the net charge and net baryon distribution realized in heavy ion experiments
the temperature and the chemical potential may be estimated at the time
of chemical freeze-out. 
}

\FullConference{31st International Symposium on Lattice Field Theory - LATTICE 2013\\
		July 29 - August 3, 2013\\
		Mainz, Germany}

\begin{document}

\section{Introduction}

Heavy ion experiments at the Large Hadron Collider (LHC) and the Relativistic
Heavy Ion Collider (RHIC) have set out the goal to produce and study new forms
of strongly interacting matter, such as the quark gluon plasma.
Besides direct emissions, we can observe this matter at the point of break-up
through the hadrons leaving the system. Prominent approaches include
the hydrodynamical modelling of the angular distribution and the study 
of the event-by-event distribution of conserved charges \cite{Hippolyte:2012yu}.

\begin{figure}
\begin{center}
\includegraphics[width=3.5in]{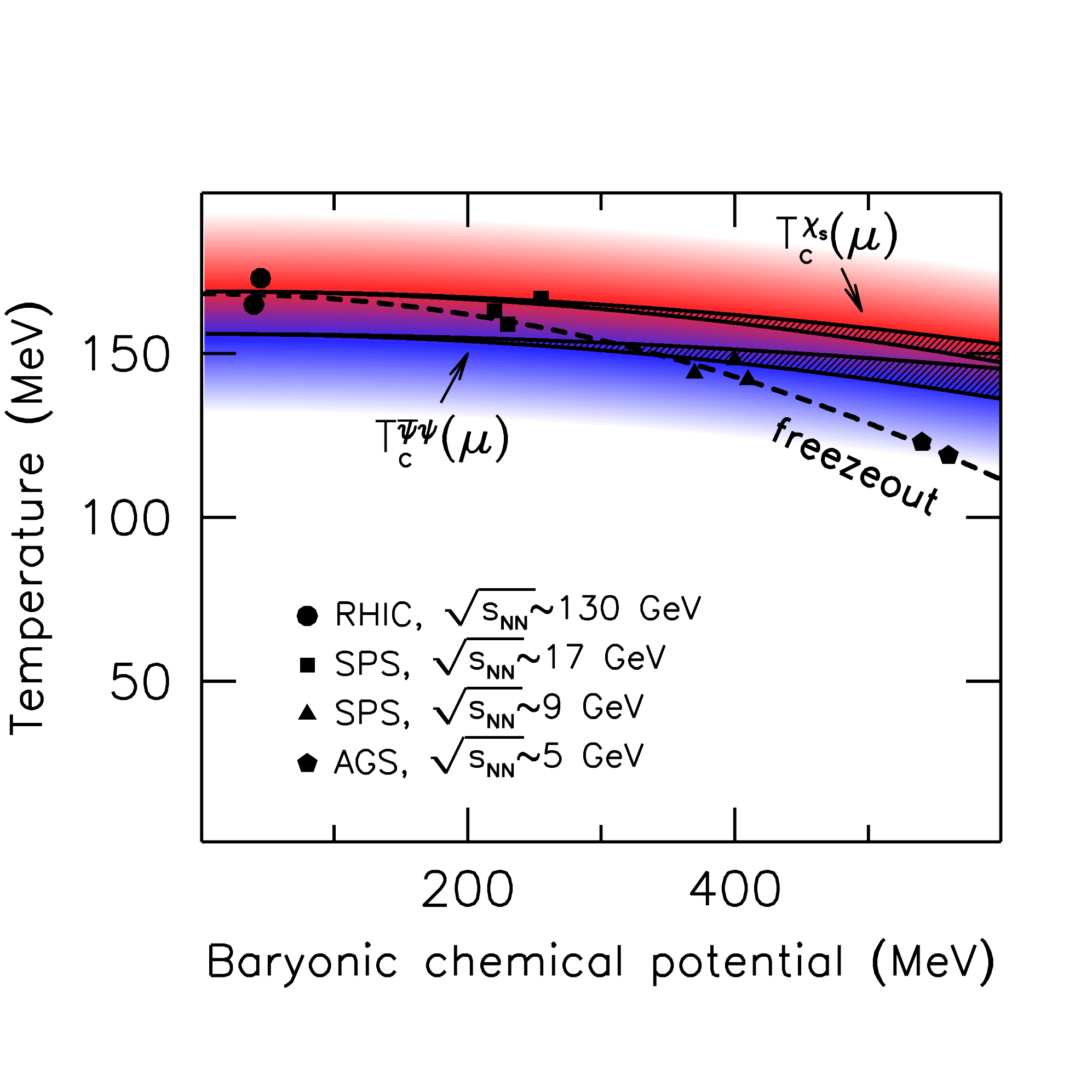}
\end{center}
\caption{\label{fig:phasediag}
The QCD phase diagram for small chemical potentials \cite{Endrodi:2011gv}.
A temperature and a chemical potential has been fitted in terms of the
statistical hadronization model for every collision energy
\cite{Andronic:2005yp,Cleymans:2005xv}. For comparison we show the crossover
lines based on two observables from lattice simulations \cite{Endrodi:2011gv}.
}
\end{figure}

The chemical freeze-out, defined as the last inelastic scattering of
hadrons before detection, has already been studied in terms of the
statistical hadronization model  by fitting a chemical potential and
a temperature parameter to the pion, kaon, proton and other accessible yields
from experiment \cite{Andronic:2005yp,Cleymans:2005xv}. For higher collision
energies smaller chemical potential are realized at freeze-out. 
Repeating the analysis for a series of beam energies provide a manifold
of $(T-\mu)$ pairs on the phase diagram, the freeze-out curve
in Fig.~\ref{fig:phasediag}.

While we know from lattice simulations that the QCD transition is a crossover
at zero chemical potential \cite{Aoki:2006we}, a critical end point and a
first order transition line may exist in the ($T$-$\mu$) plane. Its
experimental search is based on the analysis of event-by-event fluctuations
\cite{Stephanov:1999zu}.

Parallel to the experimental effort lattice field theory has been able to
describe the QCD transition in an increasing detail. The transition temperature
has been determined \cite{Tctrilogy,Bazavov:2011nk}, and the curvature of the
transition line was also given \cite{Endrodi:2011gv}. The equation of state has
been calculated at zero \cite{Borsanyi:2010cj,Borsanyi:2013bia} and small
chemical potentials \cite{Borsanyi:2012cr}. Quark number susceptibilities have
also been determined both for strange as well as light flavors
\cite{Borsanyi:2011sw,Bazavov:2012jq}. All these results have been subject to a
continuum extrapolation.

The ever-increasing accuracy of fluctuation measurements at RHIC and LHC
allows us today to make direct comparisons of lattice results with data.
The STAR experiment has recently published the beam-energy and centrality
dependence of the net-proton distribution \cite{Adamczyk:2013dal}. For
the net electric charge distribution there are preliminary results available
both from the STAR \cite{McDonald:2012ts,Sahoo:2012bs} and from the
PHENIX collaboration \cite{Mitchell:2012mx}.

The strategy for a successful comparison between theory and experiment has been
long worked on \cite{Jeon:2000wg,Asakawa:2000wh,Karsch:2012wm}. Here we use
the observables suggested in Ref.~\cite{Bazavov:2012vg}. 
The fluctuations for a conserved quantum number, such as electric charge,
are measured in a sub-system, small enough to behave like a grand canonical
ensemble, yet large enough to behave like an ensemble. The selection of a
subsystem is accomplished through cuts in rapidity and transverse momentum.
Still, the fluctuations or even the mean value of net charge depends on
the unknown subvolume. To cancel this factor ratios are considered, such
as mean/variance, which was described as a baryometer in
Refs.~\cite{Karsch:2012wm,Bazavov:2012vg}. Other relevant combinations 
are listed in Eq.~(\ref{eq:observables}).

At zero chemical potential the mean and skewness vanish, leaving
us only with the kurtosis and variance to work with at the energies of LHC.
RHIC, however, works at non-zero chemical potentials. There we expand
the lattice results around zero chemical potential and extrapolate to small but
finite values and use then the mean and the skewness, which are now non-zero.
In Ref.~\cite{Bazavov:2012vg} these observables were used as
baryometer and thermometer, respectively.

The rules for such an extrapolation are given by the experimental setting:
there is no strangeness input in the colliding nuclei, and the ratio
of protons and neutrons in the gold or lead atoms predeterimne the
charge-to-baryon ratio in the outcoming hadrons as well. Thus:
\begin{eqnarray}
\left\langle S\right\rangle=0,&&
\left\langle Q\right\rangle=0.4\left\langle B\right\rangle.
\label{eq:constraint}
\end{eqnarray}
These conditions can be respected if we introduce a strange and electric charge
chemical potential in addition to the baryochemical potential, as it has already
been a method in the statistical hadronization model.

\section{Fluctuations from the lattice}

We generated finite temperature ensembles using the three-level Symanzik
improved gauge action with dynamical stout-improved staggered fermions (see
Ref.~\cite{Aoki:2005vt}. The temporal extent of the lattices determine
the lattice spacing at a given temperature, we use
$N_t=6,~8,~10,~12,~16$ (around $T_c$ these translate to the lattice spacings of
$a=0.22, 0.16, 0.13, 0.11$ and $0.08$~fm, respectively).
At every lattice spacing and temperature we stored and analyzed every 10th
configuration in the rational hybrid Monte Carlo streams. 

\begin{figure}
\includegraphics[width=\textwidth]{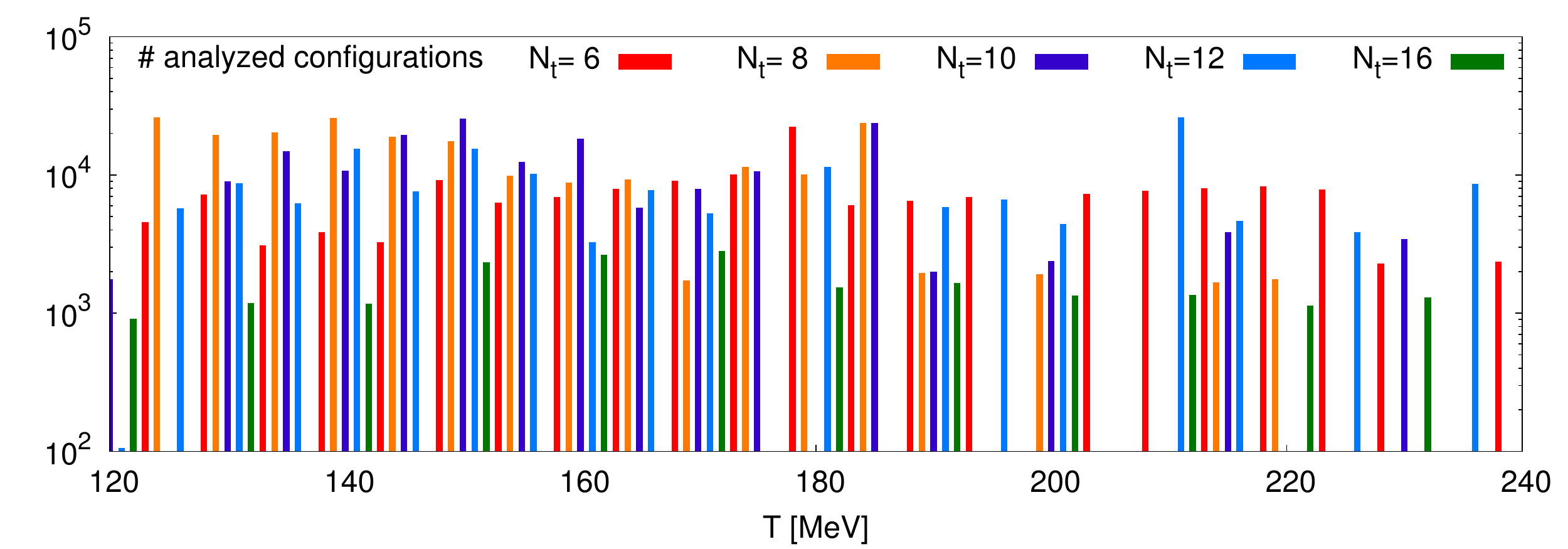}
\caption{Statistics behind the fluctuation calculations. 
The stored configurations have been separated by 10 HMC trajectories, each.
Each configuration was analyzed by $(128 \dots 256)\times 4$ random sources. \label{fig1}}
 \end{figure}

In a grand canonical ensemble we obtain the fluctuations as derivatives of the partition function with respect to the chemical potentials:
\bea
\frac{\chi_{lmn}^{BSQ}}{T^{l+m+n}}=\frac{\partial^{\,l+m+n}(p/T^4)}{\partial(\mu_{B}/T)^{l}\partial(\mu_{S}/T)^{m}\partial(\mu_{Q}/T)^{n}}.
\eea
and they are related to the moments of the distributions of the corresponding conserved charges
 by
\bea
\mathrm{ mean:}~~M=\chi_1~~&&~~\mathrm{ variance:}~~\sigma^2=\chi_2
\nonumber
\\
\mathrm{ skewness:}~~S=\chi_3/\chi_{2}^{3/2}
~~&&~~
\mathrm{kurtosis:}~~\kappa=\chi_4/\chi_{2}^{2}\,.
\label{eq:observables}
\eea
With these moments we can express the volume independent ratios
\bea
~S\sigma=\chi_3/\chi_{2}
\quad&;&\quad
\kappa\sigma^2=\chi_4/\chi_{2}\nonumber\\
M/\sigma^2=\chi_1/\chi_2
\quad&;&\quad
S\sigma^3/M=\chi_3/\chi_1\,.
\label{moments}
\eea

The chemical potential dependence enters through the fermion determinant ($\det
M_i$), allowing for one $\mu_i$ parameter for each of the three dynamical
flavor $i=u,d,s$. The actual observables are based on the derivatives of the logarithm of these determinants:
\begin{eqnarray}
A_j&=\frac{d}{d\mu_j} \log(\det M_j)^{1/4} = &\trt M_j^{-1} M_j'\,,\\
B_j&=\frac{d^2}{(d\mu_j)^2} \log(\det M_j)^{1/4} =&\trt\left(
M_j'' M_j^{-1}
-M_j' M_j^{-1} M_j' M_j^{-1}
\right)\,,\\
C_j&=\frac{d^3}{(d\mu_j)^3} \log(\det M_j)^{1/4} =&\trt\left(
M_j' M_j^{-1}
-3 M_j'' M_j^{-1} M_j' M_j^{-1}\right.\nonumber\\
&&
\left.
+2 M_j' M_j^{-1} M_j' M_j^{-1} M_j' M_j^{-1}
\right)\,,\\
D_j&=\frac{d^4}{(d\mu_j)^4} \log(\det M_j)^{1/4} = &\trt\left(
M_j'' M_j^{-1}
-4 M_j' M_j^{-1} M_j' M_j^{-1}
+12 M_j'' M_j^{-1} M_j' M_j^{-1} M_j' M_j^{-1}\right.\nonumber\\
&&
\left.
-3 M_j'' M_j^{-1} M_j'' M_j^{-1}
-6 M_j' M_j^{-1} M_j' M_j^{-1} M_j' M_j^{-1}M_j' M_j^{-1}
\right)\,.
\end{eqnarray}
We calculate these traces for every configuration using 
$(128\dots256)\times 4$ random sources. The final derivatives emerge as
connected and disconnected contributions, e.g. to second order we have
\begin{eqnarray}
\partial_i\partial_j \log Z&=&
\avr{A_iA_j}+\delta_{ij}\avr{B_i}\,.
\end{eqnarray}
Where products of diagrams appear, a disjoint set of random sources are
used, like here in $A_i$ and $A_j$, even when $i=j$. The first (disconnected)
term is responsible for most of the noise, lattice artefacts, on the other
hand, come mainly from the connected contributions.

\section{Results}

The quantities that we look at, in order to extract the freeze-out temperature
and baryon chemical potential, are the ratios 
$R_{31}^Q(T,\mu_B)=\chi_3^Q/\chi_{1}^Q$ and
$R_{12}^Q(T,\mu_B)=\chi_1^Q/\chi_{2}^Q$ for small chemical potentials,
where $\mu_Q(\mu_B)$ and $\mu_S(\mu_B)$ are chosen to satisfy
Eqs.~(\ref{eq:constraint}). We also calculated the analogous baryon
fluctuations.  For details, see the journal version of this work
\cite{Borsanyi:2013hza}.

\begin{figure}
\begin{center}
\includegraphics[width=2.9in]{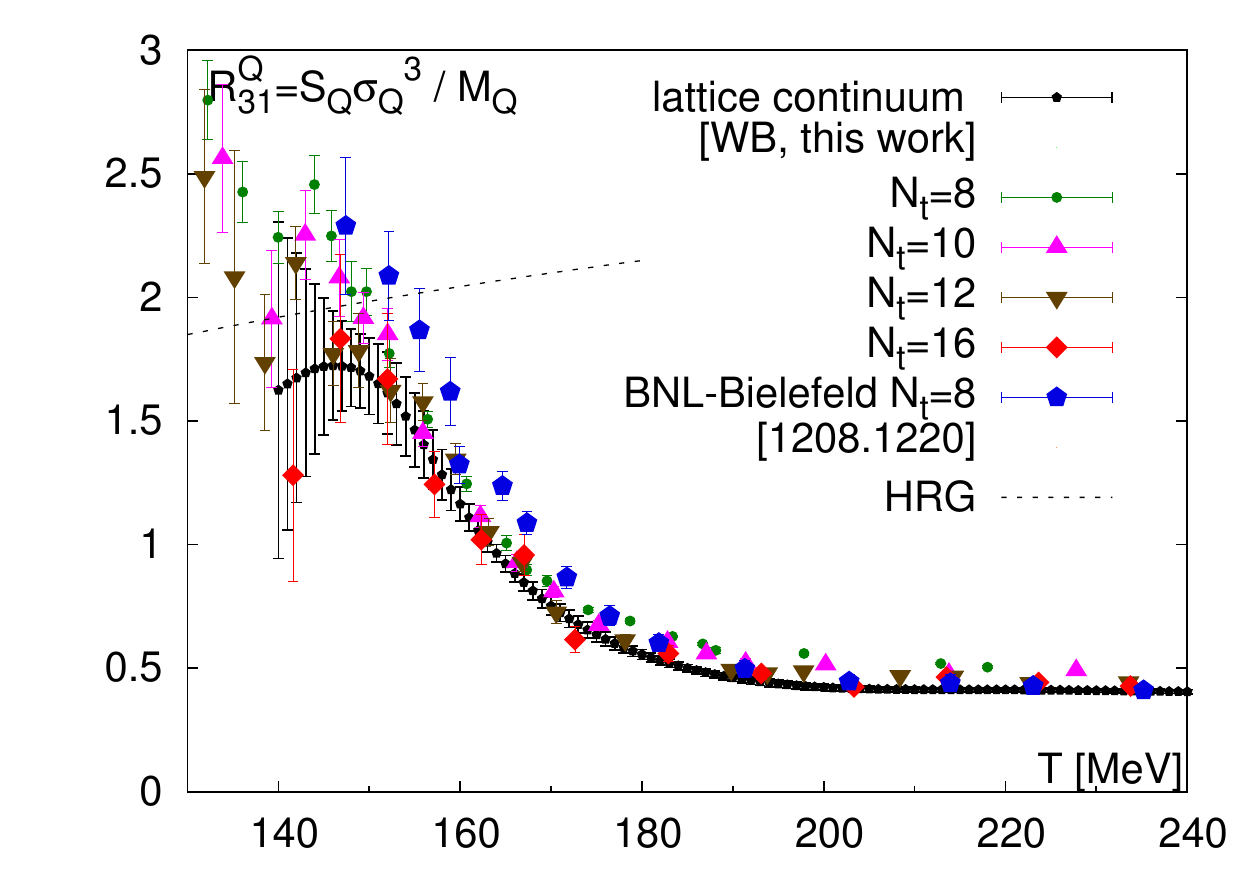}\hfil
\includegraphics[width=2.9in]{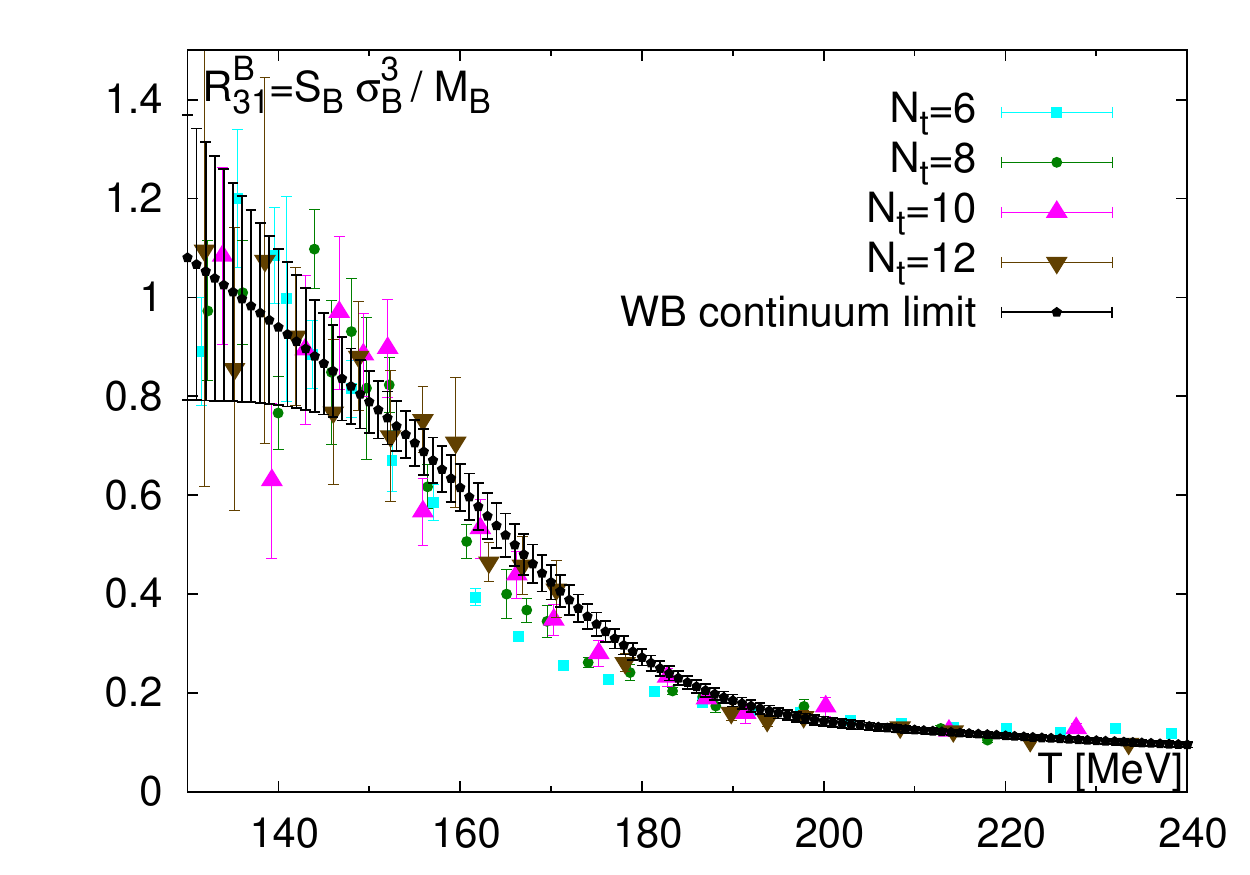}
\end{center}
 \caption{
Lattice results on the skewness ratio for the charge (left) and the baryon
number (right). The colored symbols correspond to lattice QCD simulations at
finite-$N_t$.  Black points correspond to the continuum extrapolation
\cite{Borsanyi:2013hza}; blue
pentagons are the $N_t=8$ results from the BNL-Bielefeld collaboration
\cite{Bazavov:2012vg}
\label{fig:RQ31}}
 \end{figure}

\begin{figure}
\begin{center}
\includegraphics[width=2.9in]{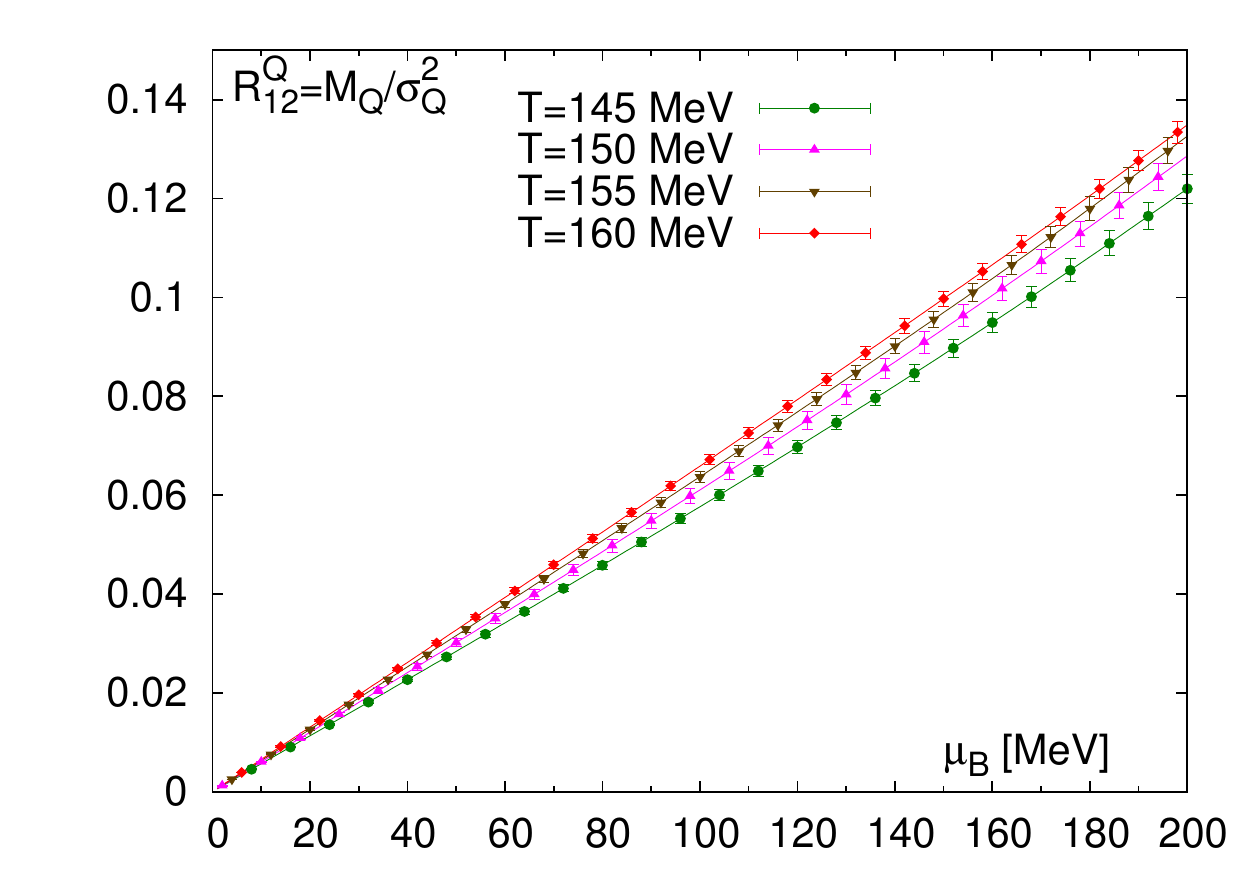}\hfil
\includegraphics[width=2.9in]{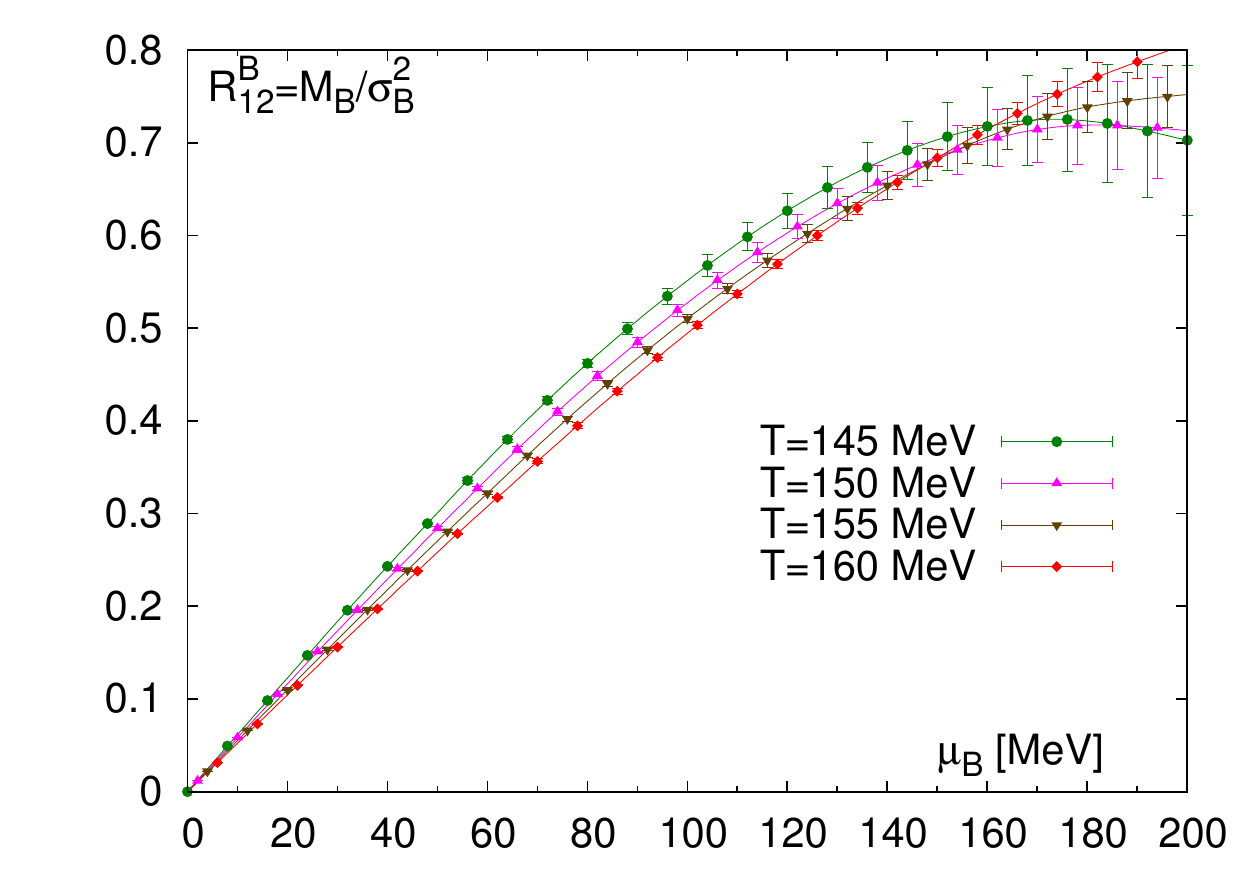}
\end{center}
\caption{$R_{12}^Q$ as a function of $\mu_B$: the different colors correspond
to the continuum extrapolated lattice QCD results, calculated in a range of
temperatures around the QCD crossover \cite{Borsanyi:2013hza}. \label{fig:R12Q}}
 \end{figure}

In Fig. \ref{fig:RQ31} we show the ratios $R_{31}^Q$ (left) and $R_{31}^B$
(right) as a function of the temperature.  The continuum extrapolations are
shown as black dots. For the charge fluctuations we used five lattice spacings.
Baryon fluctuations are plagued by greater noise, but are less sensitive to
cut-off effects, here we used four spacings.  Charge fluctuation results from
the BNL-Bielefeld collaboration corresponding to $N_t=8$ (from
Ref.~\cite{Bazavov:2012vg}) are also shown for comparison.

In Fig. \ref{fig:R12Q} we show our results for $R_{12}^Q$ as a function of the
baryon chemical potential: the different curves correspond to different
temperatures, in the range where freeze-out is expected. Such expectations
may come from the arguments in Ref.~\cite{BraunMunzinger:2003zz} supporting
a freeze-out just below the transition. Alternative hints come from the
existing estimates from the statistical hadronization model
\cite{Andronic:2005yp,Cleymans:2005xv}.  Similarly to the electric charge
fluctuations, $R_{31}^B$ will allow us to constrain the temperature and using
$R_{12}^B$ we can then obtain $\mu_B$.

Notice that the ordering of the temperatures in Fig.~\ref{fig:R12Q} (left) and
(right) is opposite. Thus, whether the chemical potentials from the charge
and the baryon (proton) fluctuations deliver consistent results will very
much depend on the associated temperature, which we can extract from the
skewness analysis. A possible source for inconsistencies might be
the comparison of proton fluctuation data with baryon fluctuations from the
lattice, and also the remnant effects of baryon number conservation
\cite{Bzdak:2012an}.  A cross-check between the freeze-out parameters from
proton and electric charge data also test the basic assumption of equilibrium
at the time of freeze-out.

\begin{figure}
\begin{center}
\includegraphics[width=2.9in]{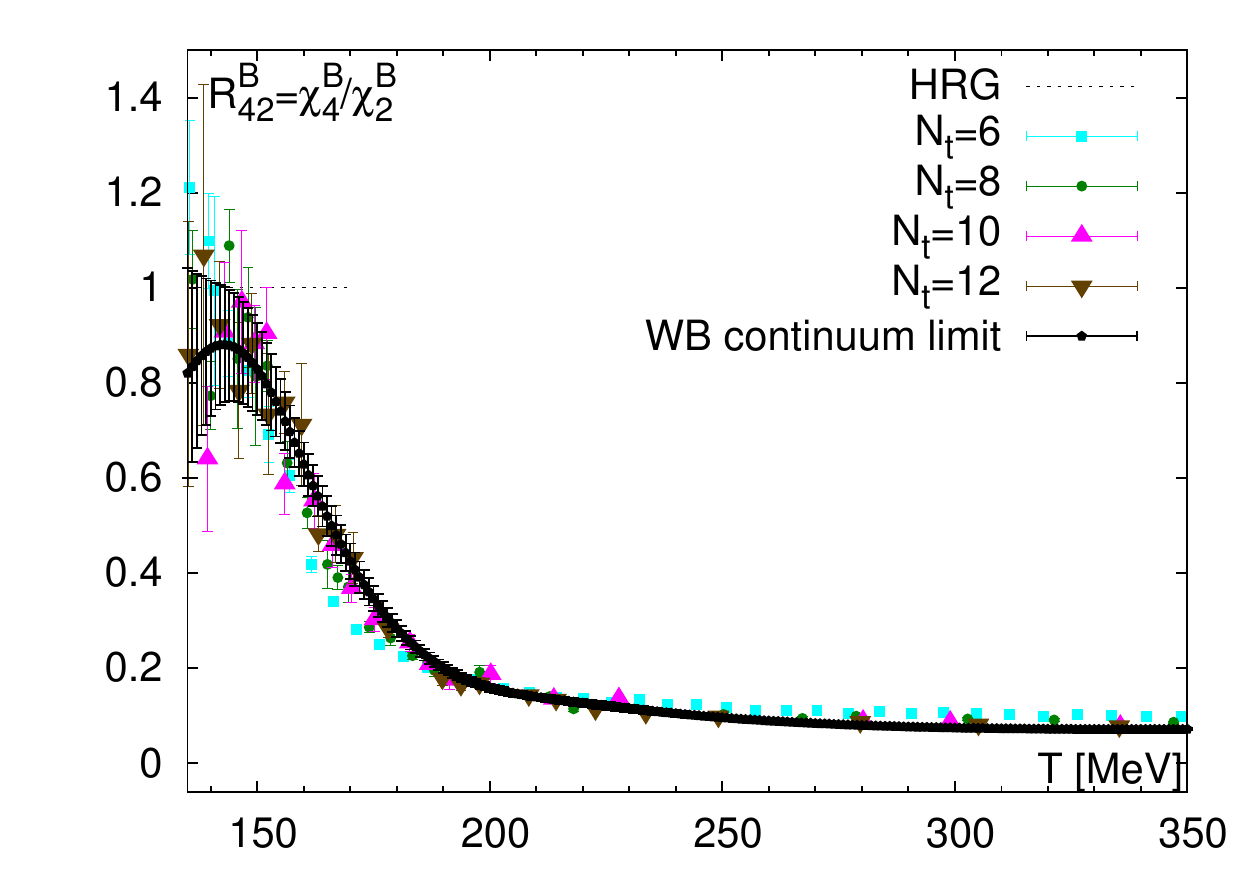}\hfil
\includegraphics[width=2.9in]{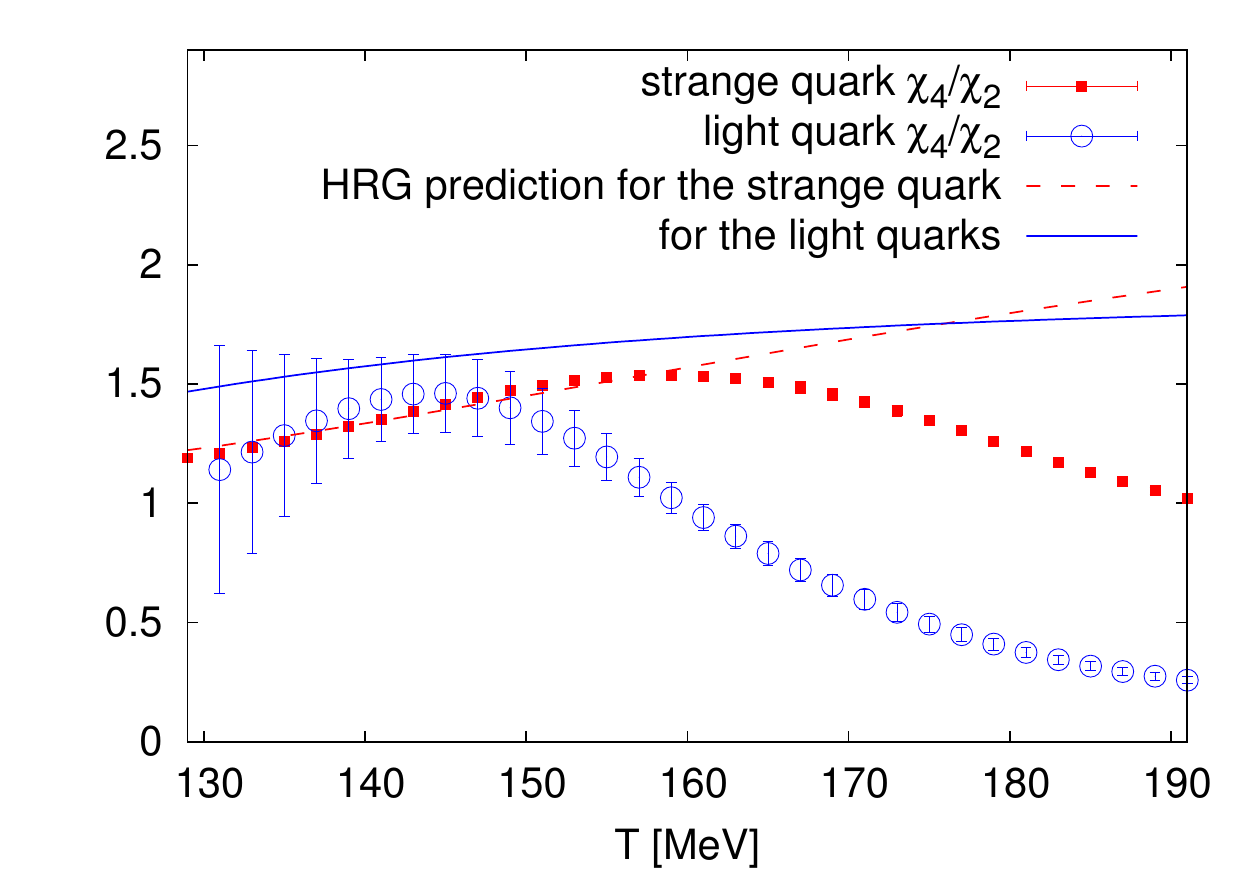}
\end{center}
\caption{\label{fig:kurtosis}
The baryon number (left) and flavor specific (right) kurtosis 
($\kappa\times \sigma^2$) prediction from lattice QCD. These parameters are
in principle accessible to LHC experiments, and may be used to define
the freeze-out temperature for specific flavors, or to the system as a whole.
}
\end{figure}

Finally we show the kurtosis data in the continuum limit in
Fig.~\ref{fig:kurtosis}. The kurtosis of baryon number and light vs. strange
quark numbers show different sensitivity to temperature, so are the maxima
and the deviation point from the hadron resonance gas prediction flavor
dependent. The great question that the experiment will have to decide is
whether the freeze-out temperatures themselves are flavor dependent
\cite{Bellwied:2013cta}.

\textbf{Acknowledgments:}
This project was funded by the DFG grant SFB/TR55.
The work of C. Ratti is supported by funds provided by the Italian Ministry of
Education, Universities and Research under the
Firb Research Grant RBFR0814TT.
S. D. Katz is funded by the ERC grant ((FP7/2007-2013)/ERC No 208740)
as well as the "Lend\"ulet" program of the Hungarian Academy of Sciences
((LP2012-44/2012).
The numerical simulations were in part performed the GPU cluster at the
Wuppertal University as well as on QPACE, funded by the DFG.  We acknowledge
PRACE  for awarding us access to the Blue Gene/Q system (JUQUEEN) at
Forschungszentrum J\"ulich, Germany.

\end{document}